# Study on the Real-time Lossless Data Compression Method Used in the Readout System for Micro-pattern Gas Detector

Zhongtao Shen, *Member, IEEE,* Shuwen Wang, Cheng Li, Changqing Feng, *Member, IEEE,* and Shubin Liu, *Member, IEEE*

*Abstract*—The data compression technology now is fully developed and widely used in many fields such as communication, multi-media, image information processing and so on. The large physical experiments, especially the ones with Micro-pattern Gas Detectors (MPGD), which always have many readout channels and have a lot of data to be transferred and saved, are however relatively seldom use this technology. In this paper, the real-time lossless data compression method is proposed for a general-purposed MPGD readout system. The lossless data compression can reduce the data transmission bandwidth of the system as well as keep all information of the data. The compression method discussed in the paper mainly consists of two steps. The first step is to pre-process the data according to different characteristics of different signals and the second step is to compress the pre-processed data using common lossless compression algorithm. Besides, the whole compression method is implemented in the Field-Programmable Gate Array (FPGA) and able to run real-timely. The system is then used to readout two different kinds of signals and the compression rate can reach as high as 43% and 30% respectively.

## I. INTRODUCTION

NOWDAYS, as the increasing requirements of physical experiments, the detectors see more readout channels and higher precision demands of signal measures [1-3]. This trend puts high pressure on the processing of data transmission and saving. However, the data compression technology, which has been used in many fields to solve the problem of data transmission and saving is almost not adopted in the field of physical experiments.

According to the theory of Claude Elwood Shannon [4], the max compression rate of data is decided by its entropy. It is known that the true random number has higher entropy then other data. The physical experiments are totally different from the random number, the signals of which always have their characteristics and the characteristics come from the physical process, the feature of detectors and readout electronics.

Therefore, the conclusion can be drawn that it is not only necessary but also suitable to used data compression technology in big physical experiments.

In this paper, the compression method is used based on a general-purposed Micro-pattern Gas Detector (MPGD) readout system [5]. The MPGD which usually has thousands of readout channels is a good test object for the data compression method. The data compression process is implemented in the Field-Programmable Gate Array (FPGA) on the MPGD readout electronics, which can run real-timely in the system.

The compression method discussed in the paper mainly consists of two steps. The first step is to pre-process the data according to different characteristics of different signals and the second step is to compress the pre-processed data using common lossless compression algorithm. Moreover, the lossless compression strategy is chosen in this paper, which means all information can be kept after compression.

Besides, the compression method is tested in two different scenarios: the general micromegas output signals and the PandaX-III output signals. Due to the different characteristics of the signals, different compression methods are proposed. And the details will be discussed in the following parts of this paper.

## II. MPGD READOUT SYSTEM

A MPGD readout system which is designed for process output signals of MPGD is chosen to test the compression method in this paper. As shown in Fig. 1, the system is split into two kinds of boards, one is for analog signal conditioning and digitization, called Front-end Card (FEC), and the other is for interfacing FECs and data processing, called data collection card (DCC). Because of the constraint of size and large number of readout channels, FEC uses custom application-specific integrated circuits (ASICs) to readout the detector signals. The ASIC chosen is a mature chip called AGET that is developed by GET collaboration targeting as a readout chip [6].

Manuscript received June 24, 2018.
This work was supported by State Key Laboratory of Particle Detection and Electronics (Grant No. SKLPDE_ZZ_201817).
(Corresponding author: Shubin Liu).

Zhongtao Shen, Shuwen Wang, Cheng Li, Changqing Feng and Shubin Liu are with State Key Laboratory of Particle Detection and Electronics, and Department of Modern Physics, University of Science and Technology of China, No.96, Jinzhai Road, Hefei, 230026, China (e-mail: liushb@ustc.edu.cn; henzt@ustc.edu.cn).

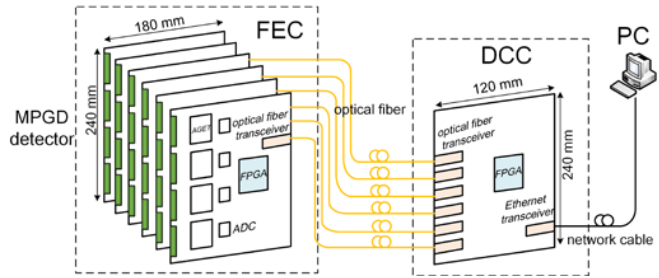

Fig. 1. The architecture of the MPGD readout system. The system is split into two kinds of boards, the FECs and the DCCs.

The AGET chip comprises 64 channels of charge sensitive amplifier with four selectable gains (120 fC, 240 fC, 1 pC and 10 pC), 16 selectable values of peaking time (50 ns to 1 us), Switched Capacitor Array (SCA) structure based 512 sample analog memory, selectable sampling rate and discriminator for trigger. And a 12bit ADC digitizes output signal of AGET on the FEC.

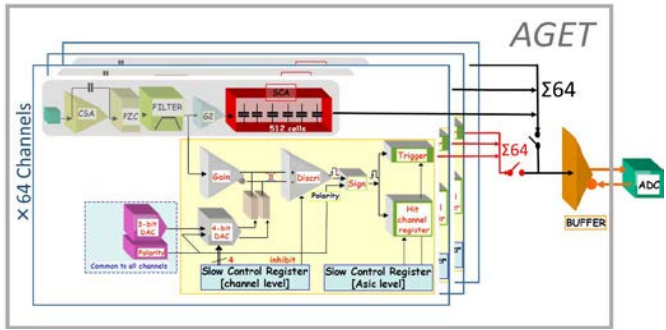

Fig. 2. Structure of the AGET chip.

The optical fiber which can transmit data at the speed as high as 1Gbps is used to link the FEC and the DCC while the DCC only adopts one network cable to send output data to PC. Besides, the system is a scalable system and the DCC can accumulate data from variable number of FECs. The input bandwidth of DCC may higher than the output bandwidth of the DCC, so it is necessary to adopt compression strategy in DCC to solve the problem.

### III. GENERAL MICROMEGAS SIGNALS

#### A. Signal Characteristics and Pre-processing

The output signal of the micromegas detector is a charge signal and this signal is sent to the front ASIC chip, the AGET chip. After the process of integration, shaping and filtering in AGET, the sampling wave is a quasi-Gaussian voltage signal, and the rising edge, the falling edge and the width of the signal are definite and decided by the analog circuits in AGET.

This is the feature of signals, based on which the pre-processing can be handled. Because of that all waves have similar shapes, if we set a reference wave and then calculate the differences between the normalized sampling waves and the reference wave, the dynamic range can be reduced. As shown in Fig. 3, the delta data are much smaller than the raw data and the 12-bit original data can be compressed to 8-bit delta data.

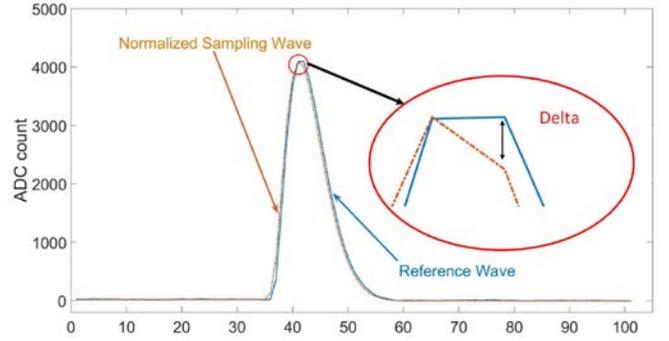

Fig. 3. Reference wave and sampling Wave of AGET.

#### B. Common Compressing Algorithm

After the first step, the 12-bit raw data have been compressed to the 8-bit data. Indeed, the purpose of pre-processing is not to compress the data but to transform the raw data to achieve higher compression rate in the next step. Generally, common lossless compression method includes Huffman coding, LZ coding, Run-length coding and based on the feature of the data to be compressed, one of the methods can be chosen.

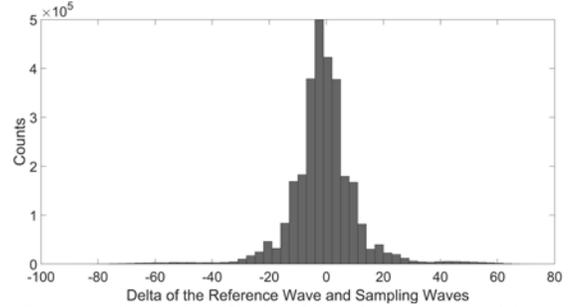

Fig. 4. Frequency distribution of Delta of the reference wave and the sampling waves. The frequencies data appear are highly relative to the data values. The data closer to zero appear more frequently.

By counting the delta data, the feature of them can be reflected. Fig. 4 is the frequency distribution of delta data after pre-process discussed before. It is obvious that the number closer to zero appears more frequently and this is because the reference wave comes from many sampling waves and reflects the feature of the circuits in AGET.

The Huffman Coding [7] which encodes based on the exact frequencies is then chosen as the second step compression algorithm for two reasons. Firstly, the delta data have the feature that the frequencies data appears are high related to the data values. And the second reason is that every data package is small and different data package is relatively unrelated to each other. Therefore, the compression methods based on dictionary such as LZ coding are abandoned.

Besides, since every data package has the same statistics characteristics as shown in Fig. 4, the static Huffman coding table is used to simplify the design and the number closer to zero is replaced by shorter code.

#### C. FPGA Implementation

As mentioned before, the data compression process is planed to be implemented in FPGA to make sure that it can run real-timely. The data flow in FPGA of the compression procedure is shown in Fig. 5. The data coming into the FPGA are first cached in a FIFO, and then the Peak Searching Module search for the peak location and the peak value of data in FIFO. By using the searching results, the raw data can be normalized, aligned and compared with the reference wave saved in the FPGA before. And finally, the delta data are sent to the Huffman core to code based on a static table saved in FPGA before.

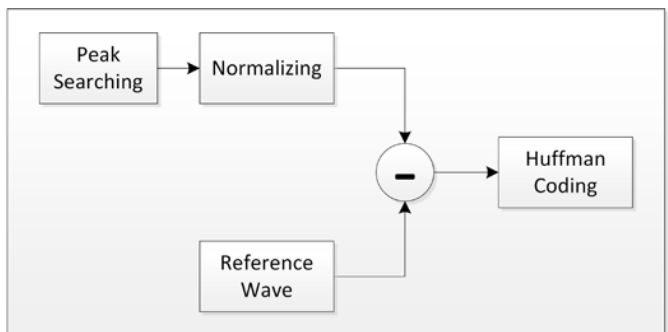

Fig. 5. Compression processing.

In the Huffman core, the coding process runs according to a static table and a complicated combinational logic is implemented to accomplish this function. The combinational logic which consists of large numbers of look-up tables and gate circuits cannot work at a very high speed. To solve this problem, a parallel structure is adopted, as shown in Fig. 6. In this structure, many Huffman cores work simultaneously at a relatively low speed but the whole module can deal with data at a very high speed, which can guarantee the real-time performance of the system.

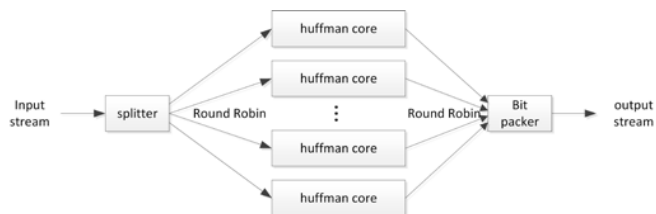

Fig. 6. The structure of parallel Huffman module. This structure is to solve the problem that the Huffman core can only encode at a relatively low speed.

### D. Tests and Results

The MPGD readout system with data compression method discussed before is then under a joint test with a micromegas detector and the test data are analyzed. After the pre-process, the compression rate can achieve 67% and after two steps, the compression rate of 43% is got.

## IV. PANDAX-III EXPERIMENTS

### A. Signal Characteristics and Pre-processing

PandaX-III is an experiment which uses a radio-pure high-pressure gaseous xenon TPC to search for the Neutrinoless Double Beta Decay (NLDBD) of 136Xe under the circumstance of 200kg, 90% 136Xe at 10bar.

Different from normal the TPC signals, the output signals of PandaX-III may have widths of more than 10us, which are longer than the integration time of the AGET because of the long trace of NLDBD event. Compare to the quasi-Gaussian wave, the AGET output signals of PandaX-III may have a long flat top, as shown in Fig. 7. The signals of the event can be obtained by calculating the areas of the waves instead of measuring the peaks value of the waves [8].

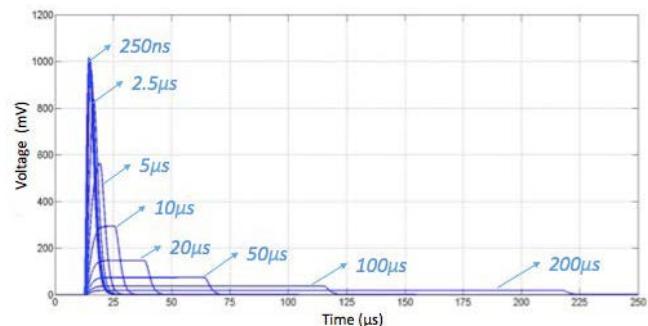

Fig. 7. Simulation results of AGET output signals of PandaX-III.

By observing the signals, it can be found that except for the rising and falling edges, most parts of the waves almost have a slope of zero. Therefore, if calculating the delta between next two points, we can get the results most of which are around zeros and this strategy is used to pre-process the raw data of PandaX-III experiments.

### B. Common Compressing Algorithm

If counting the data after pre-process, the feature of them can be found. As shown in Fig. 8, most data fall near the zero point. As result, the Huffman coding is again chosen as the compression algorithm.

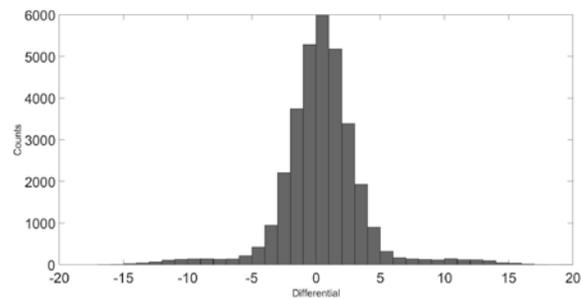

Fig. 8. Frequency distribution of the differential of the wave. The frequencies data appear are highly relative to the data values. The data closer to zero appear more frequently.

### C. FPGA Implementation

The data flow in FPGA is shown in Fig. 9. The implement logic mainly consists of two parts, the pre-processing part and the Huffman core. The purpose of pre-processing part is to calculate the differential of the raw data. There are two data paths in the FPGA and one of them has a one clock delay. By calculating the difference of the data from two paths, the differential of raw data is got. The Huffman core, which accomplish data compression, uses the same logic module as the one mentioned in section III.

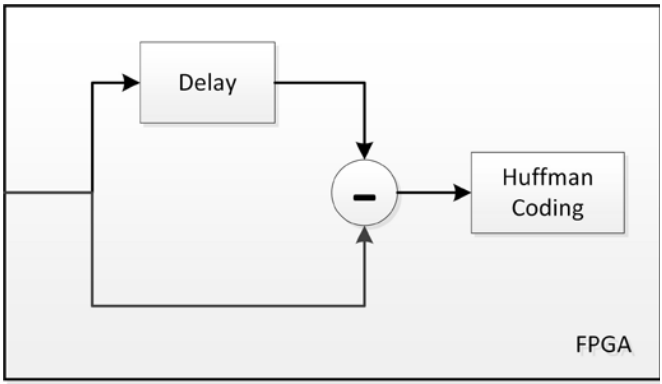

Fig. 9. Compression processing.

### D. Tests and Results

Because the NLDBD case is very rare, the signal generator is used to simulation the signals and the data compression rate is tested on this condition. The test site is shown in Fig. 10. The signal generator AFG3252 from Tektronix generates a step voltage signal and this signal is then injected into an adapt board. There is a capacity on the adapt board which can transform a voltage signal into a charge signal. By controlling the rising edge of the voltage signal the width of the charge signal can be controlled. The charge signal which is regarded as the simulation of the NLDBD signal is then sent to the MPGD readout system. By analyze the test results, the compression rate in this situation is obtained and it reaches as high as 30%.

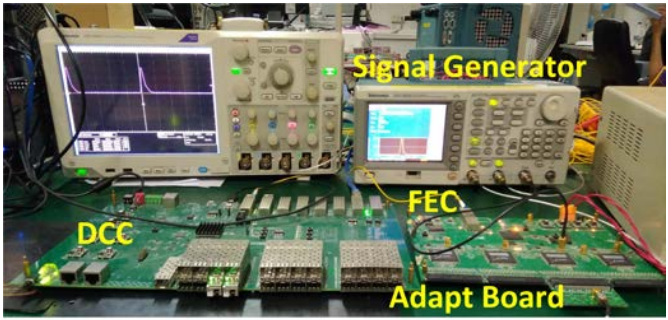

Fig. 10. Test Site.

## V. Discussion

### A. Performance and Resource Consumption

The FPGA used in the DCC is xc7z045ffg676-2 from Xilinx company. The resource consumptions of data compression strategy in the two situations discussed before are shown in Table I. From the table it can be seen that in both situations the compression logics only consume very few hardware resources. Besides, by analyzing the implementation tools and simulation tools of the FPGA the system time constrains can be obtained. In both situations, the data compression logics can run well under the system clock of 100MHz. Therefore, the conclusion can be drawn that the data compression strategy has the advantages of few resource consumption and good real-time performance, which may lead to a good application prospect.

TABLE I
RESOURCE CONSUMPTION

|  | Slice LUTs | Slice Registers | Block RAM |
|---|---|---|---|
| Situation1 | 438 (0.20%) | 372 (0.09%) | 5 (0.92%) |
| Situation2 | 394 (0.18%) | 293 (0.07%) | 1 (0.18%) |

### B. More Applications

In this paper, facing two different scenes two different data compression strategies are adopted and the strategies are both based on a MPGD readout system. However, this does not mean that the data compression is only suitable to MPGD. In fact, all other detectors which meet pressure of data transmission and saving can try to adopt the data compression strategy with the advantages of few resource consumption and good real-time performance. The only difference is to design different compression methods according to different features of the output signals.

## VI. Conclusion

In this paper, the data compression method is used in MPGD readout system to solve the problem in data transmission. The system is tested in two different experiments and both get a good compression rate. Besides MPGD, this method can be used in other detector readout system in the future.